\documentstyle[preprint,tighten,aps,psfig]{revtex}

\begin{document}

\preprint{hep-ph/9706311, TPI-MINN-97/19, TTP97-13}
\title{Complete \boldmath${\cal O}(\alpha _s ^2)$ Corrections to 
Zero-Recoil Sum
Rules for $B \to D^*$ Transitions} 
\author{Andrzej Czarnecki\thanks{Address after September 1997: 
Physics Department,
Brookhaven National Laboratory,  Upton, New York 11973.} 
and Kirill Melnikov}
\address{Institut f\"ur Theoretische Teilchenphysik, \\
Universit\"at Karlsruhe,
D-76128 Karlsruhe, Germany}
\author{Nikolai~Uraltsev\thanks{Permanent address:
Petersburg Nuclear Physics Institute,
Gatchina, St.~Petersburg 188350, Russia}}
\address{Theoretical Physics Institute, University of Minnesota,\\
Minneapolis, MN 55455, USA}
\maketitle

\begin{abstract}
We present the complete ${\cal O}(\alpha _s ^2)$ corrections to the
Wilson coefficient of the unit operator in the zero-recoil sum rule
for the $B \to D^*$ transition.  We include both perturbative and
power-suppressed nonperturbative effects in a manner consistent with
the operator product expansion.  The impact of these corrections on
$|V_{cb}|$ extracted from semileptonic $B\rightarrow D^*$ decays near
zero recoil is discussed.  The mixing of the heavy quark kinetic
operator with the unit operator at the two loop level is
obtained. ${\cal O}(\alpha_s)$ corrections to a number of
power-suppressed operators are calculated.
\end{abstract}
\vspace*{2cm}

\newpage
\section{Introduction}
Semileptonic decays of $B$ mesons provide an opportunity to measure
the Cabibbo-Kobayashi-Maskawa matrix parameter $|V_{cb}|$ with minimal
theoretical uncertainties (for a recent review, see
e.g. \cite{rev}). One of the two most popular methods is based on the
experimental determination of the zero-recoil $B\rightarrow D^*$ transition
amplitude by extrapolating the experimental decay rate of $B\rightarrow
D^*\,\ell \nu$ to the point of zero recoil momentum, where the
invariant mass squared of leptons is $q^2_{\ell \nu} =
(M_B-M_{D^*})^2$. The hadronic $B\rightarrow D^*$ transition amplitude for
this kinematics is written as:
\begin{equation}
\langle{D^*}|{\bar{c} \gamma_\mu \gamma_5 b}|{B}\rangle
 \;=\; 2\, F_{D^*} \sqrt{M_B
M_{D^*}}\; e^*_\mu\;,
\label{1}
\end{equation}
where  
$e$ is the polarization vector of the $D^*$ meson.
The zero-recoil form factor $F_{D^*}$ 
is calculable in the short-distance perturbative
expansion up to terms $\sim (\Lambda_{\rm QCD}/m_{c,b})^2$. These 
nonperturbative
corrections cannot  be evaluated in a model-independent way at present;
they are expected to be about $-8\%$ \cite{rev}.

Existing estimates of the long-distance strong interaction corrections
to $F_{D^*}$ are based on the sum rules for heavy flavor transitions
\cite{vcb,optical}. They relate certain sums of the transition
probabilities to expectation values of local heavy quark operators in
the decaying hadron. The zero-recoil sum rule for the spatial
components of the axial current can be written in the form:
\begin{equation}
| F_{D^*}|^2 + \sum_{\epsilon<\mu}\,|F_{\rm exc}|^2 \;=\;  
\xi_A(\mu)\;-\; \frac{\xi_\pi(\mu)}{m_c^2} \mu_\pi^2(\mu) \;-\;
\frac{\xi_G(\mu)}{m_c^2} \mu_G^2(\mu) \;+\; {\cal
O}\left(\frac{\mu^3}{m_Q^3}\right)\;.
\label{5}
\end{equation}
The functions $\xi_A$, $\xi_\pi$ and $\xi_G$ are 
short-distance (perturbative) 
coefficient functions, $m_Q$ denotes heavy quark masses $m_{b,c}$ and 
$\mu_\pi^2$, $\mu_G^2$ are $B$-meson expectation values of the kinetic
and chromomagnetic operators, respectively:
\begin{equation}
\mu_\pi^2(\mu) = \frac{1}{2M_{B}} \langle{B}|{\bar b
(i\vec{D})^{\,2} b}|{B}\rangle_\mu\;,\;\;\;\;
\mu_G^2(\mu)  = \frac{1}{2M_{B}} \langle{B}|{\bar b \frac{i}{2}
\sigma_{\alpha\beta} G^{\alpha\beta} b}|{B}\rangle_\mu\;.
\label{6}
\end{equation}
By $F_{\rm exc}\,$ we generically denote transition form factors
between 
$B$ and
excited charm states with masses $M_D+\epsilon$. They are related to the
appropriate structure function of the $B$ meson:
\begin{eqnarray}
| F_{D^*}|^2 + \sum_{\epsilon<\mu}\,|F_{\rm exc}|^2 &=&  
\frac{1}{2\pi}\, 
\int_0^\mu w_1^A(\epsilon\,,\vec{q}=0) \,{\rm d}\epsilon\;,
\label{7}
\\[1.5mm]
w_1^A &=& \frac{2}{3}\,{\rm Im}\, h_{ii}^A\qquad \epsilon=
M_B-M_{D^*}-q_0\;\;,
\end{eqnarray}
with the invariant hadronic
amplitudes $h$ defined as in \cite{optical}:
\begin{eqnarray}
\hat T_{\mu\nu} &=& i\int {\rm d}^4x {\rm e}\,^{-iqx}\:
{\rm T}\left\{ j^\dagger _\mu(x)\, j_\nu(0)\right\} \qquad 
(\mbox {here }
j_\mu = \bar c \gamma_\mu \gamma_5 b),
\nonumber\\[1.5mm]
h^A_{\mu\nu} &\equiv & {1\over 2M_B}\langle B|\hat T_{\mu\nu}|B\rangle.
\end{eqnarray}
The derivation of such  
sum rules and their usage for estimating physical form factors 
 is explained in detail in \cite{optical}.

Because of short-distance perturbative effects the sum over excited
states in the l.h.s.\ of the sum rules does not converge at $\epsilon
\sim \Lambda_{\rm QCD}$. Instead, for $\Lambda_{\rm QCD} < \epsilon <
m_Q $ one has $w_1^A(\epsilon\,,\vec{q}=0)\sim \alpha_s (\epsilon)
\epsilon/m_Q^2$.  This necessitates introducing a cutoff at some
energy $\mu$. The cutoff serves as a natural normalization point of
the effective low-energy operators, most appropriate for use with the
sum rules. The scale $\mu$ satisfies the constraint $\Lambda_{\rm QCD}
\ll \mu \ll m_{c,b}$.\footnote{In reality, this amounts to use $\mu
\sim$ several units times $\Lambda _{\rm QCD}$. The existence of such
a scale in practice is the criterion for applicability of the heavy
quark expansion to charm quarks at a quantitative level.}  As
explicitly indicated in Eq.~(\ref{5}), all coefficient functions are
also $\mu$-dependent.

The leading term $\xi_A(\mu)$ is of primary importance, since it
accounts for the short-distance perturbative renormalization of the
zero-recoil axial current. It is calculable in the perturbative
expansion provided $\mu$ is large enough and belongs to the
perturbative domain.  If $\mu$ is too large, it weakens the
constraining power of the sum rules, since the heavy quark expansion
runs in powers of $\mu/m_Q$.

The ${\cal O}(\alpha_s )$ perturbative corrections to the  
sum rules were
calculated in \cite{optical} (see also \cite {kmy}). Separate pieces
of BLM--corrections \cite{blm} contributing to sum rules at order
${\cal O}(\beta _0 \alpha_s  ^2)$ were considered in
\cite{neub,bbbsl,lig}. The complete BLM resummation of the unit
operator coefficient function $\xi _A(\mu)$ was carried out in
\cite{blmope}. The impact of these quasi--one-loop corrections on the
sum rules proved to be small when one follows the Wilson approach to
the Operator Product Expansion (OPE) assuming explicit separation of
short-distance and long-distance contributions.

More challenging are the genuine, non-BLM ${\cal O}(\alpha_s ^2)$
corrections. Their magnitude is crucial for estimating the actual
impact of unknown higher-order corrections. Technically the most
complicated piece corresponding to virtual corrections to heavy quark
currents at zero recoil was calculated to order ${\cal O}(\alpha_s
^2)$ in \cite{czarm}. In the present paper the remaining second-order
corrections to the zero-recoil sum rule for spatial components of the
axial current are computed and the complete two-loop expression for
the perturbative coefficient function $\xi_A(\mu)$ is given.  We also
derive the two-loop evolution of the kinetic operator.  As a byproduct
of the analysis, we find ${\cal O}(\alpha_s )$ corrections to
coefficients of some power suppressed operators in the nonrelativistic
expansion, and obtain a similar correction to the coefficient of the
kinetic operator in the sum rule.  The ${\cal O}(\alpha _s ^2)$
correction to the sum rule for the timelike component of the vector
current ($B\rightarrow D$ transition) will be given elsewhere.

\section{ \boldmath${\cal O}(\alpha_s^2)$ corrections to $\xi _A(\mu)$}
The most efficient way to determine $\xi_A(\mu)$ was suggested in
\cite{optical} (see also \cite{upset,blmope}). It relies on
considering the OPE relations of the type of Eq.~(\ref{5}) in
perturbation theory. The structure function $w_A$ is then given by the
weak transitions amplitudes between initial and final states
consisting of quarks and gluons.  Our aim here is to calculate it to
order ${\cal O}(\alpha_s ^2)$.

The l.h.s.\ has an elastic contribution $b\rightarrow c$ for on-shell
quarks and the continuum contribution from $b\rightarrow c+{\rm
gluon}$ and $b\rightarrow c+{\rm 2\:gluons}$. The elastic contribution
is equal to $|\eta_A^{(2\,\rm loop)}|^2$, which was calculated in
\cite{czarm}. The inelastic part of the structure function to order
${\cal O}(\alpha_s ^2)$ is calculated in the present paper as an
expansion in $\mu/m_Q$.  In order to determine perturbative
coefficients in the sum rules, we need the inelastic part only through
the leading, second order in $\mu/m_Q$, because as far as
nonperturbative effects emerging from local higher-dimension operators
are concerned, we account explicitly only for $1/m_Q^2$
terms.\footnote{The $1/m_Q^3$ corrections were also calculated
\cite{rev}. In this paper, however, we limit our consideration to the
leading $1/m_Q^2$ nonperturbative corrections.}

Therefore, in order to calculate $\xi_A(\mu)$ with ${\cal O}(\alpha _s
^2)$ accuracy, one has to calculate $\eta_A^2$, the  
inelastic part of the
structure function $w_1^A$, the  perturbative correction to the coefficient
function $\xi_\pi(\mu)$ of the kinetic operator and the  perturbative
expectation value of the kinetic operator to the  necessary order in
$\alpha_s $. The expectation value of the chromomagnetic operator vanishes
in the perturbative expansion to leading order in $1/m_Q$.

Through order $\alpha_s ^2 \mu^2/m_Q^2$  the perturbative contribution 
to the l.h.s.\ of Eqs.~(\ref{5}), (\ref{7}) has the form 
(the Feynman diagrams for the inelastic part are shown in Fig.~1):
\begin{equation}
\frac{1}{2\pi}\,
\int_0^\mu w_1^{A\,({\rm pert})}(\epsilon\,,0) \,{\rm d}\epsilon\;
= \;\left(\eta_A^{(\rm 2\,loop)}\right)^2 + 
\left \{\frac {\alpha_s (M)}{\pi}
C_F \Delta_1^A 
+  \Bigg (\frac {\alpha_s}{\pi} \Bigg )^2 C_F \Delta _2^A (\mu;\,M)
\right \} \left(\frac {\mu}{m_c} \right)^2 ,
\label{8}
\end{equation}
where $M$ is a  normalization point for the strong coupling 
constant
in the $\overline{\rm MS}$ scheme.
Denoting   $x = m_c/m_b$, we get:
\begin{equation}
\Delta_1^A  = \frac {1}{4} 
\Bigg (1+
\frac {2}{3}x +x^2 \Bigg ),
\end{equation}
\begin{equation}
\Delta _2^A = C_F\Delta _F^A + C_A\Delta _A^A + T_R N_L \Delta _L^A,
\end{equation}
\begin{eqnarray}
 \Delta _F^A (\mu) &=&
          \ln\left( x\right)   \left(
          -  { 2 \over 1-x }
        +   {13 \over 8}
          + x
          + {3 \over 8} x^2
          \right)
          - {25 \over 36}
          - {11 \over 18} x
          - {25 \over 36} x^2
,
\\[5mm]
\Delta _A^A (\mu) &=&
  \ln\left( x\right)   \left(
          - {1 \over 4}
          - {1 \over 12} x
          \right)
          + \Delta _1^A
\left[\ln \left(\frac {2\mu}{m_c} \right) 
 - {5\over 24}\pi^2 + {11\over 6} \ln \left(\frac{M}{2\mu}\right) \right] 
\nonumber \\ &&
\quad
           + {917 \over 720}
          + {949 \over 1080} x
          + {917 \over 720} x^2
,
\\[5mm]
\Delta _L^A (\mu) &=&
          - {77 \over 180}
          - {89 \over 270} x
          - {77 \over 180} x^2
     - \frac {2}{3}\, \Delta _1 ^A \ln \left(\frac{M}{2\mu}\right).
\end{eqnarray}
\vspace*{0.2cm}

Although the sum of the elastic $b\rightarrow c$ transition probability and
the inelastic excitations is more infrared-stable than they are
separately, it is still not completely free of power suppressed
infrared contributions.  Its infrared part is given by the expectation
values of the local operators in the r.h.s.\ of the sum rule
(\ref{5}).  Therefore, by calculating the expectation values of the
local operators in perturbation theory and accounting for them in
Eq.(\ref {5}), we eliminate  the contribution of the
infrared domain from the Wilson coefficient of the unit operator $\xi
_A(\mu)$.

The expectation value of the kinetic operator in perturbation
theory can be determined by using the sum rule for
spatial components of the vector current at 
zero recoil  in the heavy quark limit \cite{vcb,optical,blmope}:
\begin{equation}
\frac{1}{2\pi}\,
\int_0^\mu w_1^V(\epsilon) \,{\rm d}\epsilon\;= \;
\frac{\xi^V_\pi}{m_c^2} \mu_\pi^2(\mu) \;-\;
\frac{\xi^V_G(\mu)}{m_c^2} \mu_G^2(\mu) \;\;.
\label{13}
\end{equation}
The excitation probability (the perturbative structure function) 
$w_1^V(\epsilon)$
is calculated using the same technique as for the axial current. To apply 
this sum rule, however, we need the coefficient functions
$\xi^V_{\pi}$ 
with ${\cal O}(\alpha_s )$ accuracy. For this purpose
we perform a 
nonrelativistic
expansion of the vector current accounting for the ${\cal O}(\alpha _s)$ 
corrections. The result of this calculation reads:
\begin{eqnarray}
\bar{c} \vec\gamma b\;\vert_{\vec{q}=0} &=& 
\varphi_c^+ \left( -u\,i\vec{D} \;+\; v\, [\vec\sigma \times \vec D]
\right) 
\varphi_b\;,
\label{11}
\\[1.5mm]
u&=& {1\over m_b}\left\{
 {1+x\over 2x}
\left [1-C_F\frac{\alpha_s }{\pi}\left(\frac{3(1+x)}{4(1-x)} 
\ln (x)
+1 \right)
\right]\; + \; 
 C_F\frac{\alpha_s }{2\pi}\frac{\ln(x)}{1-x}\;\right\},
\nonumber \\
v &=& {1\over m_b}
 {1-x\over 2x}
\left [1-C_F\frac{\alpha_s }{\pi}\left(\frac{3(1+x)}{4(1-x)} 
\ln (x)
+1 \right)
\right].
\nonumber 
\end{eqnarray}
This is the ${\cal O}(\alpha_s )$--corrected form of the 
expansion for the vector current
given in \cite{optical}, Eq.~(181).  The coefficients $u$ and $v$ are
obtained by evaluating one-loop graphs shown in Fig.~2 in the
linear approximation
in
$\vec{p}_{b,c}/m_Q$. Using Eqs.~(\ref{11}) one 
determines the normalization of the  state produced by the vector
current: 
\begin{equation}
\frac{1}{3} \sum_n |\langle{n}|{\bar{c}\gamma_i b}|{B}\rangle|^2 \;=\; 
\frac{1}{3}\left [ 
(u^2+2v^2)\, \mu_\pi^2 \;-\;(v^2-2uv)\, \mu_G^2 
\right ]\;.
\label{12}
\end{equation}
As was shown   in Ref. \cite {optical},
this normalization is nothing but the sum rule of interest. 
We get
\begin{eqnarray}
\xi^V_\pi &=& 
\frac{1}{4}\left\{
\left(1 -\frac{2}{3}x + x^2 \right)
-
2C_F\frac{\alpha_s }{\pi}
\left[ \left( 1 -\frac{2}{3}x + x^2
\right) +\frac{3}{4}\frac{1+x}{1-x}\ln{x}
\left(1 - \frac{10}{9}x +x^2
\right)
\right]
\right\},
\label{xipiv}
\\
\xi^V_G &=&
\frac{1}{4}\left\{
\left(-\frac{1}{3}-\frac{2}{3}x + x^2 \right)
-
2 C_F \frac{\alpha_s }{\pi}
\left[\left(-\frac{1}{3}-\frac{2}{3}x + x^2 \right) 
 -\ln x
\left(\frac{1}{4} + \frac{2}{3}x + \frac{3}{4}x^2
\right)
\right]
\right\}\,.
\label{xigv}
\end{eqnarray}
The same expression for $\xi_\pi^V$ can be also obtained 
by considering the sum rule for a slowly moving $b$ quark.
Below, this method is applied to derive 
the Wilson coefficient 
of the kinetic operator
$\xi_\pi$ which enters the axial sum rule 
[cf. Eqs.~(\ref{16})--(\ref{16b})].

In order to get the expectation value of the kinetic operator in
perturbation theory and, therefore, its mixing with the unit operator
to ${\cal O}(\alpha_s ^2)$ accuracy, one has to evaluate the l.h.s.\
of the sum rule (\ref{13}) to second order in perturbation theory  
through 
terms $\mu ^2/m_Q^2$.  Since the chromomagnetic operator does not
mix with the unit operator, the perturbative contribution as a whole
should be identified with the kinetic operator.

Performing this calculation, we get, 
with the ${\cal O}(\alpha _s ^2)$ accuracy:
\begin{equation}
\left(\mu_\pi^2(\mu)\right)_{\rm pert}
  \;=\;C_F\,\frac{\alpha_s (2\mu)}{\pi}\: \mu^2  
\;+\; C_F\,\left(\frac{\alpha_s }{\pi}\right)^2\,\left[ 
\left(\frac {91}{18} -\frac {\pi ^2}{6}\right) C_A - 
\frac{13}{9}T_R N_L
\right]\:\mu^2\;,
\label{15a}
\end{equation}
\begin{equation}
\frac{{\rm d}\mu_\pi^2(\mu)}{{\rm d}\mu^2} = 
C_F \frac{\alpha_s (\mu)}{\pi} +
\left(\frac{5}{3} -\ln{2} \right)
 C_F \left( \frac{11}{6}C_A-\frac{2}{3} T_R N_L \right)
\left(\frac{\alpha_s }{\pi}\right)^2  +
\left( \frac{13}{12}-\frac{\pi^2}{6} \right) 
C_A C_F \left(\frac{\alpha_s }{\pi}\right)^2.
\label{15}
\end{equation}
The last term in Eq.~(\ref{15}) 
represents the non-BLM  contribution. Note that the 
term $\sim C_F^2$ is absent. 
It means 
that
the first-order mixing in the
Abelian theory without light flavors is not renormalized by 
effects of higher orders. 
We note that this fact actually holds   to all orders in 
perturbation theory (see Ref. \cite{blmope}).

Finally, we calculate the coefficient function $\xi _{\pi}(\mu)$
of the kinetic operator in the axial
sum rule Eq.~(\ref{5}) to order  ${\cal O}(\alpha_s )$. 
The result is as follows: 
\begin{eqnarray}
\xi_\pi(\mu) &=& \xi _\pi^{(0)}\,+\,C_F \frac {\alpha_s }{\pi} \xi
_\pi^{(1)}(\mu)\,+\,{\cal O}\left(\frac{\mu}{m_Q}\alpha_s ,\,\alpha_s
^2\right)\:, 
\qquad\qquad
\label{16}
\\[1.5mm]
\xi _\pi ^{(0)} &=& \frac{1}{4} 
\left(1+\frac{2}{3}x + x^2\right),
\nonumber \\
\xi _\pi^{(1)}(\mu) &=& \frac {2}{3}(1-x)^2\ln\left( \frac
{2\mu}{m_c}\right) 
+\frac {9-17x+31x^2-7x^3}{24(1-x)}\ln x +\frac {1+22x+x^2}{18}.
\label{16b}
\end{eqnarray}
The last relation is obtained by 
considering the zero-recoil sum rule perturbatively, in the
first order in $\alpha_s $ for  the initial $b$ quark moving with 
spatial momentum $|\vec{p}\,| \ll m_b,~m_c$. 
Let us note that for $m_b \ne m_c$, the zero recoil condition 
$\vec{q}=0$
implies a change in the spatial quark velocity.
This change leads to 
gluon bremsstrahlung. The virtual corrections become infrared
divergent, this divergence, as usual, is 
compensated by the contribution of the  real gluon emission. 
This cancelation, however, brings in an
explicit logarithmic dependence of $\xi_\pi ^{(1)}$ on $\mu$.

We note that this logarithmic dependence is not quite usual.  Although
the kinetic operator has a vanishing anomalous dimension, its
coefficient in the sum rule contains the logarithm of the cut-off
parameter $\mu$ at order $\alpha_s $. Because of  the power mixing of the
kinetic operator with the unit operator, the coefficient of the latter
has a similar logarithm in the power suppressed term
$\alpha_s ^2\mu^2/m_Q^2$.  We must emphasize, however, that the $\log \mu$
in Eq.~(\ref{16b}) does not originate from the dependence of the
coefficient function on the normalization point used for the kinetic
operator, but rather from the explicit dependence of the observable we
consider (axial sum rule) on $\mu$.  If we introduce a normalization
point $\nu$ of the operator as an independent parameter not equal to
$\mu$, there will be no $\log \nu$ dependence in $\xi_{\pi}^1$ and
only $\alpha_s  \log \mu$ will remain.  A transparent physical picture
underlying this fact will be discussed in a separate publication.

Equations (\ref{8}), (\ref{15a}) and (\ref{16})--(\ref{16b}) combined
with the known result for $\eta_A^{(2\,\rm loop)}$ allow us to obtain
$\xi_A$ to order $\alpha_s ^2$. We note, however, that our perturbative
expressions were given in terms of the pole masses of the heavy quarks
as they appear in this order of perturbation theory. To get rid of
spurious $1/m_Q$ infrared effects associated with the pole masses, we
have to switch to the short-distance masses $m_{b,c}(\mu)$ which have
concrete numerical values at  a given $\mu$, independent of the order of
perturbation theory. To order $1/m_Q^2$ this change affects only the
term $\eta_A^2$. Unless this is done, the sum rules are formally
inconsistent since the perturbative expansion of $\eta_A$ has an
infrared piece of the order of $\alpha_s (m_Q) \Lambda_{\rm QCD}/m_Q$ 
\cite{upset}.

In principle, the short-distance masses 
can be defined in different ways. Since
throughout this paper the renormalization scheme with the cutoff over the
excitation energy is implemented, 
we use the ${\cal O}(\alpha_s )$ relation \cite{five} 
\begin{equation}
\frac{{\rm d}m_Q(\mu)}{{\rm d}\mu}\;=\;- C_F\frac{\alpha_s
  }{\pi}\left(\frac{4}{3}+ 
\frac{\mu}{m_Q}+{\cal O}\left(\frac{\mu^2}{m_Q^2}\right)\,
\right)
\;.
\label{20}
\end{equation}
Then we replace
\begin{eqnarray}
\lefteqn{
|\eta_A^{(2\,{\rm loop})}\left(m_Q^{\rm pole}\right)|^2 \; \rightarrow \; 
|\eta_A^{(2\,{\rm loop})}\left(m_Q(\mu)\right)|^2 }
\nonumber \\
&& \qquad \qquad -\;
2 C_F^2
\left(\frac{\alpha_s }{\pi}\right)^2\,
\left(\frac {2x}{1-x}\ln x + 1+x \right)\;
\left( \frac {\mu}{m_c} + \frac {3}{8} (1+x) \left(\frac
{\mu}{m_c}\right)^2 \right)
\;.
\label{21}
\end{eqnarray}

The final result for the coefficient function 
$\xi _A(\mu)$
of the unit operator is obtained by combining all terms 
calculated  above. 
Following \cite{comment}, we express the result in terms of 
$\alpha_s \left(\sqrt{m_cm_b}\right)$, even though
the $m_c \leftrightarrow m_b$ symmetry arguments (which 
motivate this choice) 
do not apply for 
$\xi_A(\mu)$ because an additional momentum scale is present. 

Collecting all pieces together
and neglecting terms
$\mu^3/m_Q^3$, we obtain for $\xi_A(\mu)$:
\begin{eqnarray}
\xi_A^{(2\,\rm{loop})} (\mu) &=& \left(\eta_A^{(2\,\rm{loop})} 
(m_Q(\mu))\right)^2 +
C_F\frac {\alpha_s }{\pi}\xi _A ^{(1)}\left(\frac {\mu}{m_c}\right)^2
\nonumber \\
&& 
+\; 
C_F^2\left(\frac {\alpha_s }{\pi}\right)^2 \frac
{\mu}{m_c} \varsigma_A \;+\;
C_F\left(\frac {\alpha_s }{\pi}\right)^2 
\left(\frac {\mu}{m_c}\right)^2 \zeta_A^{(2)}\;,
\\[1.5mm]
\varsigma_A &=& -2\left( \frac {2x}{1-x}\ln x + 1+x \right),
 \\
\zeta_A ^{(2)} &=& C_F \zeta_A^F + C_A \zeta_A^A + T_R N_L \zeta_A^L,
 \\[1.5mm]
\zeta_A^F &=& \frac {2}{3}(1-x)^2 \ln \left(\frac {2\mu}{m_c} \right)
-\frac {x(17+5x+4x^2)}{6(1-x)}\ln x -\frac {25}{18} - \frac {8}{9}x
- \frac {25}{18}x^2, 
\\
\zeta_A^A &=& -\left(1+\frac {2x}{3}+x^2\right)\left[\frac {2}{3}
\ln \left(\frac {2\mu}{m_c} \right)+\frac {3\pi^2}{32}\right]\;
\nonumber \\
&& -\;
\left( \frac {17}{24} + \frac {7x}{18}+\frac {11x^2}{24} \right)\ln x
+\frac {203}{80}+\frac {1859}{1080}x+\frac {203}{80}x^2,
\\
\zeta_A^L&=& \frac{1}{3}\left(1+\frac {2x}{3}+x^2 \right) 
\left[
\ln \left(\frac {2\mu}{m_c} \right)
+{1\over 2}\ln x
\right]
 -\frac {71}{90} -\frac {77x}{135}-
\frac {71x^2}{90}.
\end{eqnarray}
The function  $\eta_A^{(2\,\rm{loop})}$ 
to order ${\cal O}(\alpha_s ^2)$ can be found in \cite{czarm}.

Since the BLM part of the corrections was discussed in the
literature in detail \cite {blmope}, 
we single out the genuine non-BLM part $a_2^0$ which
is defined as the value of the full second-order coefficient $a_2$ at
$N_L=\frac{11}{4} \frac{C_A}{T_R}\:$:
\begin{eqnarray}
\xi_A(\mu)&=&1 \,+\, a_1\left(\mu,\,m_Q(\mu)\right) \frac{\alpha_s
  }{\pi}\; +\;  
a_2\left(\mu,\,m_Q(\mu)\right) \left(\frac{\alpha_s }{\pi}\right)^2 \; +\;...
\nonumber \\
&=&\; 1 + a_1\left(\mu,\,m_Q(\mu)\right) \frac{\alpha_s }{\pi}\;+
\left[c_2^{\rm BLM}\left(\mu,m_Q(\mu)\right) 
\beta _0 +
a_2^0\left(\mu,m_Q(\mu)\right)
\right]\left(\frac{\alpha_s }{\pi}\right)^2  + ... \;,
\label{30}
\\[1.5mm]
\beta _0 &=& \frac{11}{3}C_A-\frac{4}{3}T_R N_L\;.
\end{eqnarray}
Note that $a_2^0$ does not depend on the convention 
for the normalization point
of the strong coupling. Its value is shown in Fig.~3 as a function
of $\mu/m_c$ for three 
values of $m_c/m_b \; = \; 0.2,\; 0.25$ and $0.3$. We denote
the  corresponding  non-BLM second-order shift in $\xi_A(\mu)$ by
$\Delta^0(\mu)$:
$$
\Delta^0(\mu)\;=\; a_2^0\left(\mu,\,m_Q(\mu)\right)
\left(\frac{\alpha_s }{\pi}\right)^2 \;.
$$

\section{\boldmath$|V_{cb}|$ determination and zero--recoil sum rules}
Let us now turn to the application of our results for the
determination of $|V_{cb}|$.
First,  we note that
the net impact of the non-BLM $\alpha_s ^2$ corrections on the sum rules is
rather small. Taking a reasonable value $\mu/m_c=0.5$ and $m_c/m_b=0.25$
we get $a_2^0 \approx -0.7$. Assuming $\alpha_s =0.22$ to $0.27$, the absolute
shift $\Delta^0(\mu)$ 
constitutes $-0.003$ to $-0.005$ which translates into $-0.0015$ to
$-0.0025$ decrease of the short-distance perturbative renormalization of
the zero-recoil axial current. For any reasonable choice of the
parameter $\mu$ in the sum rules consistent with using the $1/m_c$ 
expansion, this effect is well below 1\%. It
should be noted that since in the Wilson OPE the infrared domain is
completely excluded from the coefficient functions, 
the effective coupling  cannot become large.

The perturbative correction to the coefficient of the kinetic operator
(Eq.~(\ref{16b})) is strongly suppressed. The actual value of this
coefficient changes only by several percent.  We did not calculate the
corresponding effect in the chromomagnetic term; due to  the anomalous
dimension of this operator it depends on  the specific renormalization
procedure. In view of the result for the vector sum rule (see
Eq.~(\ref{xigv})) we do not expect this correction to be significant
either.

The axial sum rule (\ref{5}) allows one to get a QCD-based estimate of
the combined effect of  the perturbative and nonperturbative effects on
the $B\rightarrow D^*$ zero-recoil form factor:
\begin{equation}
|F_{D^*}|=\left[\xi_A(\mu)- \frac{\xi_\pi(\mu)}{m_c^2} \mu_\pi^2(\mu)
- \frac{\xi_G(\mu)}{m_c^2} \mu_G^2(\mu) -
\sum_{\epsilon<\mu}\,|F_{\rm exc}|^2\right]^{1/2} + \; 
{\cal O}\left(\frac{\mu^3}{m_Q^3}\right)\;.
\label{sr}
\end{equation}
It is seen that
$\sqrt {\xi_A(\mu)}$ plays the role of a short-distance renormalization
factor of the weak current, while the remaining terms in square
brackets yield $1/m^2$ long-distance corrections to the form factor. The
quantum-mechanical meaning of each of the three power terms, confirming
this formal conclusion, was elucidated in \cite{optical} (see also
\cite{rev}).

With the good perturbative control over the short-distance part of
$F_{D^*}$, the main uncertainty dominating theoretical
predictions for the form factor resides in  
power corrections.

In Ref. \cite{blmope} higher order BLM--type corrections to the sum
rule were analyzed.  It was shown that, assuming that the running of
the QCD coupling $\alpha_s$ below the charm mass is a valid practical
concept, it is necessary to perform a resummation of the leading BLM
corrections in order to arrive at a meaningful numerical
result. Within the Wilson approach to OPE, the overall impact of the
BLM corrections appears to be moderate.  The typical BLM-resummed
value for $\xi_A(\mu)$ at $\mu/m_c\simeq 0.5$ appears to be near
$0.99$.

On the practical side, excluding the infrared domain notably improves
the accuracy of the perturbative calculations for the charm quark, at
least in the context of the BLM calculus. We recall that the
perturbative zero-recoil
factor $\eta_A$
has an intrinsic uncertainty 
$\sim (\Lambda_{\rm QCD}/m_c)^2$ due to infrared
renormalons. By calculating $|\xi_A(\mu)|^{1/2}$ and taking into account 
$\mu^2/m_Q^2$ terms, the $(\Lambda_{\rm QCD}/m_c)^2$ uncertainty 
$\delta _{\rm IR}^{1/m^2}$
is eliminated. As long as we do not account for $1/m_Q^3$ terms
explicitly, the perturbative expressions still have an 
infrared renormalon uncertainty
of the order of  $(\Lambda_{\rm QCD}/m_c)^3$.  Using the same 
overall normalization for both cases one has
\begin{equation}
\delta_{\rm IR}^{1/m^2}(\eta_A) \: \sim \: (1+x)^2
\left(\frac{\Lambda_{\rm QCD}}{m_c}\right)^2
\; \rightarrow  
\;\: 
\delta_{\rm IR}^{1/m^3} \: \sim \:
\frac{3{\rm e}^{5/6}}{16}
(11+5x+5x^2+11x^3)\left(\frac{\Lambda_{\rm QCD}}{m_c}\right)^3.
\label{ir}
\end{equation}
Performing a simple estimate, one finds that
the size of the $1/m_c^2$ uncertainty in $\eta_A$ becomes quite 
significant for $\Lambda_{\rm QCD} \simeq 250\,\mbox{MeV}$. This shows
up in the BLM 
corrections which become quite large already at the lowest orders. 
Shifting the uncertainty down 
to $\sim (\Lambda_{\rm QCD}/m_c)^3$, we significantly reduce it.

The analysis of the ${\cal O}(\alpha _s ^2)$ corrections 
presented in this paper shows that the genuine two-loop 
effects are quite small.
Therefore, our numerical conclusions do not differ in practice  
from the estimate of the zero-recoil axial form factor
given   in \cite{rev}
\begin{equation}
F_{D^*}\;\simeq \; 0.91 \;-\;0.013\,
\frac{\mu_\pi^2-0.5\,\mbox{GeV}^2}{0.1\,\mbox{GeV}^2}\;\pm\;
0.02_{\rm excit}\;\pm\;0.01_{\rm pert}\;\pm\;0.025_{1/m^3}\;\;.
\label{F}
\end{equation}
For further improvements, one has to bring in a new dynamical 
input yielding the magnitude of  
long-distance $1/m_c^2$ and $1/m_c^3$ 
corrections more precisely.

Experimental data on the small-recoil $B\rightarrow D^*$ decay rate
are not fully conclusive yet. Nevertheless, the value of $|V_{cb}|$
extracted from the exclusive $B \rightarrow D^*$ transitions using
$F_{D^*} \simeq 0.9$ seems to be in a reasonable agreement --- already
within experimental uncertainties --- with the results obtained from
inclusive semileptonic decay width $\Gamma_{\rm sl}(B \to X_c l
\nu_l)$.  The estimate of the complete ${\cal O}(\alpha_s ^2)$
correction in $\Gamma_{\rm sl}(b\rightarrow X_c l \nu _l)$, recently
presented in Ref. \cite{czarmel}, is another example of the
theoretical progress in the perturbative treatment of one of the most
important heavy quark decays.
\vspace*{.2cm}

\section{Conclusions}
We have calculated the complete ${\cal O}(\alpha_s ^2)$ corrections to
the zero-recoil heavy quark axial sum rule which is used to evaluate
the zero-recoil $B\rightarrow D^*$ form factor.  The calculations
performed here are a necessary supplement to the results of
Ref. \cite{czarm}.  The use of the sum rules allows one to incorporate
both ${\cal O}(\alpha_s ^2)$ and power-suppressed nonperturbative
effects in a consistent way.  The genuine (non--BLM) two-loop
corrections are shown to be relatively small and under good
theoretical control. We note that the uncertainty associated with
these effects had not been reliably estimated previously.  This
theoretical uncertainty is now eliminated.

As an important theoretical conclusion, we emphasize that a consistent
implementation of the Wilson OPE separating short- and long-distance
contributions, is feasible even when highly nontrivial complete ${\cal
O}(\alpha_s ^2)$ corrections are taken into consideration.  In principle,
this procedure does not bring in additional complications, compared to
purely perturbative calculations performed without an infrared
cutoff. The Wilson approach, on the other hand, allows one to operate
with well defined notions of short--distance and long--distance
effects.  From the practical viewpoint, it allows one to decrease
significantly the uncertainty in perturbative coefficients in beauty
decays. This essential reduction originates mainly from the BLM-type
corrections; the genuine two-loop effects are less radically changed.

We note also, that only in the framework of the Wilson OPE approach to
QCD is it possible to preserve a number of exact inequalities
formulated for hadrons containing a heavy quark.  They exist in a
simplified quantum mechanical treatment which ignores the
peculiarities of the field-theoretical description.  These
inequalities are important in constraining those parameters of the
heavy quark expansion which are not yet measured in experiment.

As a byproduct of our analysis, we obtained the complete two-loop
perturbative evolution of the kinetic operator, 
Eq.~(\ref{15}).\footnote{The BLM part was analyzed to all orders in
\cite{blmope}.}

We also calculated ${\cal O}(\alpha_s )$ corrections to the coefficient
function of the kinetic  operator in the axial sum rule.  The
nonrelativistic expansion of $\bar{c} \vec \gamma b$ at zero recoil
and the corresponding sum rule were obtained with ${\cal O}(\alpha_s )$
accuracy.  Numerically, we found an overall short-distance
renormalization of the zero-recoil current to be very small, close to
the estimates of \cite{vcb,optical,rev} and rather different from the
value used in other analyses of $F_{D^*}$ where long-distance $1/m^2$
corrections were addressed.

\section*{Acknowledgments}
N.U. gratefully acknowledges discussions of the OPE subtleties with the
TPI fellows M.~Shifman, M.~Voloshin and A.~Vainshtein, and
valuable perturbative insights from Yu.~Dokshitzer. He also thanks I.~Bigi
for encouraging interest and R.~Dikeman for comments.  
This work was supported in part by DOE under the
grant number DE-FG02-94ER40823, by  BMBF under grant number 
BMBF-057KA92P, and by
Graduiertenkolleg ``Teilchenphysik'' at the University of Karlsruhe.

\vspace*{10mm}
\begin{figure}[h]
\hspace*{-5mm}
\begin{minipage}{16.cm}
\[
\mbox{
\hspace*{10mm}
\begin{tabular}{cc}
\hspace*{10mm}
\psfig{figure=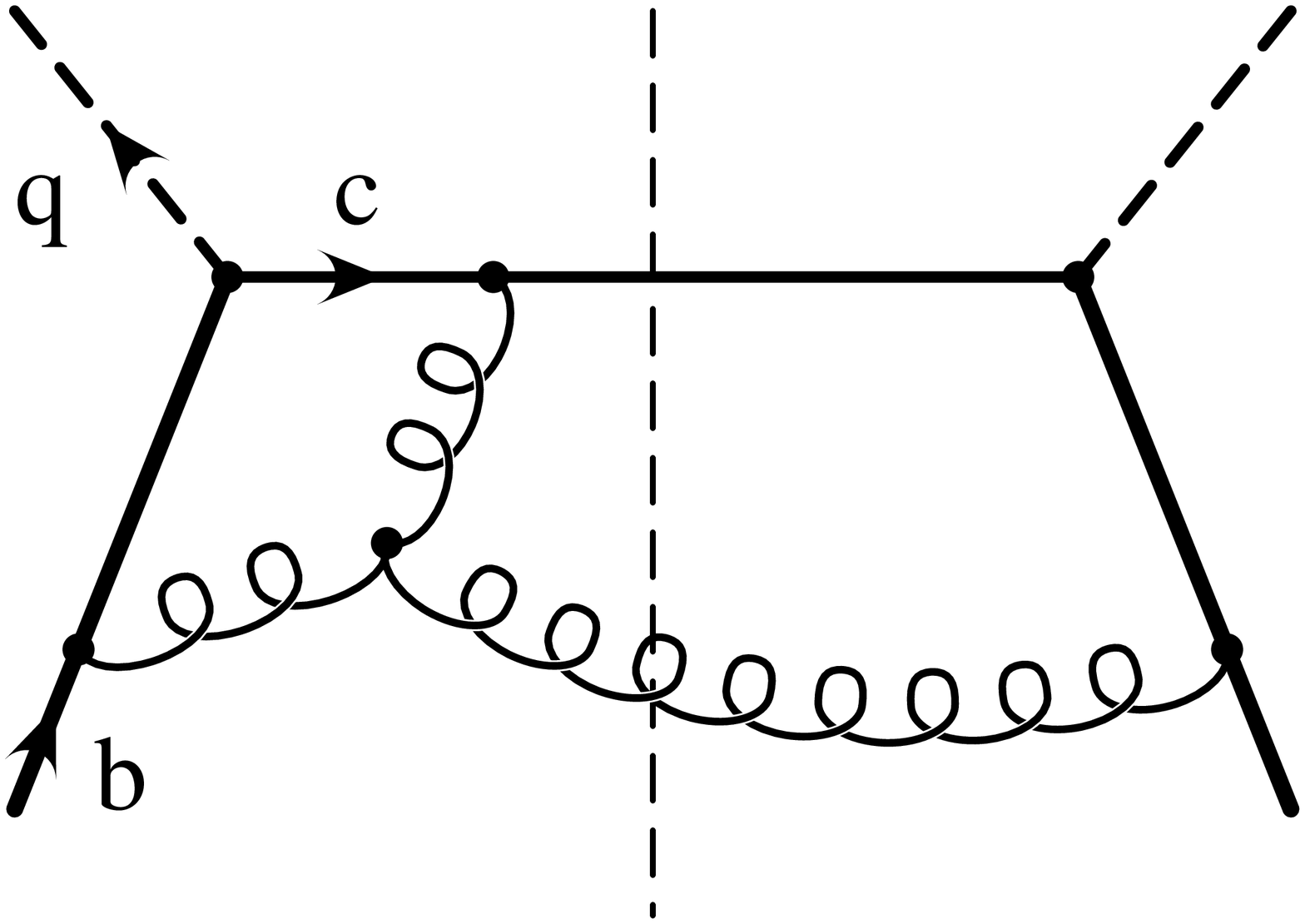,width=50mm,bbllx=210pt,bblly=410pt,%
bburx=630pt,bbury=550pt} 
&\hspace*{25mm}
\psfig{figure=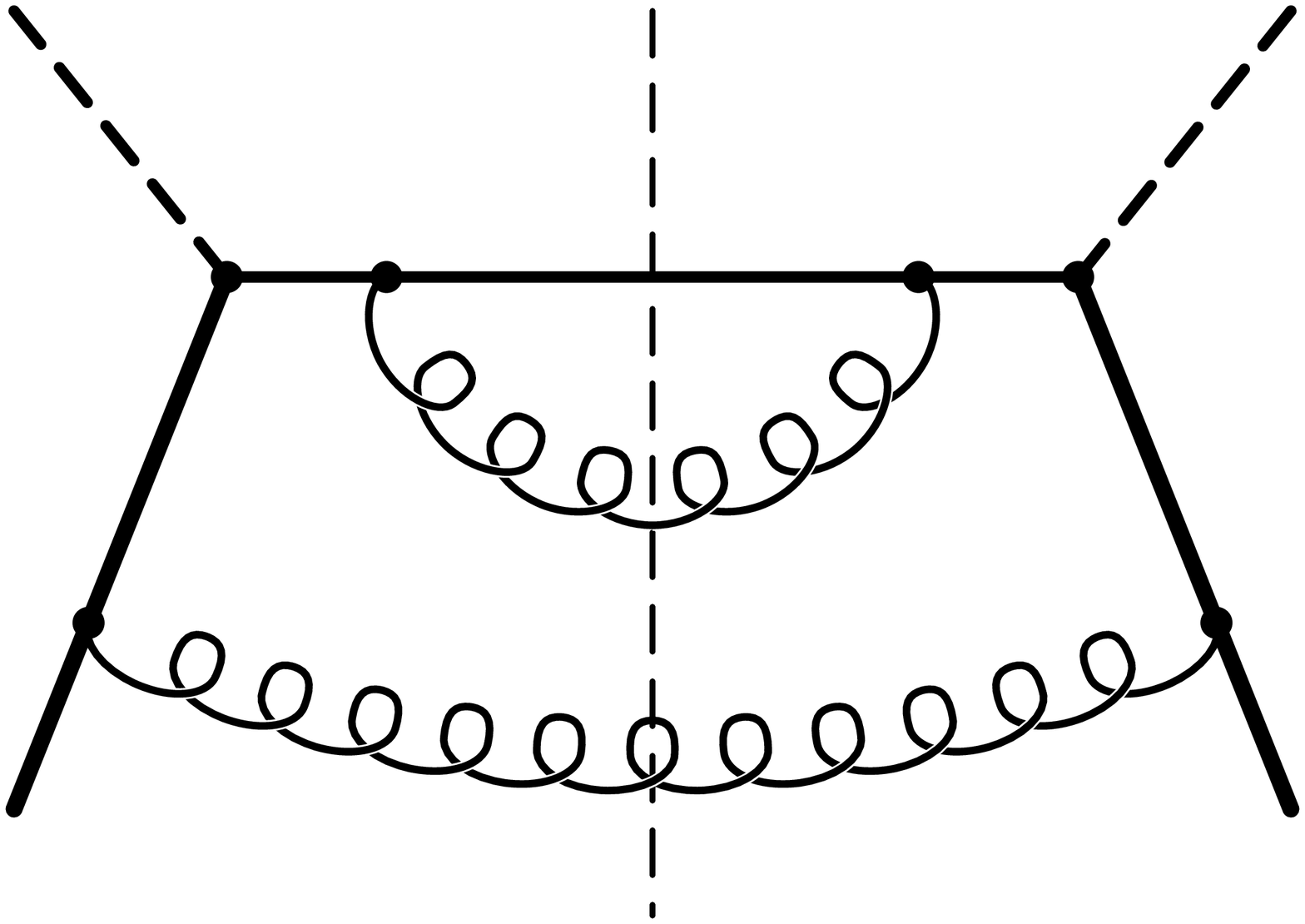,width=50mm,bbllx=210pt,bblly=410pt,%
bburx=630pt,bbury=550pt}
\\[40mm]
\end{tabular}}
\]
\end{minipage}
\caption{Examples of inelastic contributions to the zero recoil
  sum rules.} 
\label{fig:inel}
\end{figure}

\vspace*{10mm}
\begin{figure}[h]
\hspace*{-5mm}
\begin{minipage}{16.cm}
\[
\mbox{
\hspace*{10mm}
\begin{tabular}{ccc}
\hspace*{-2mm}
\psfig{figure=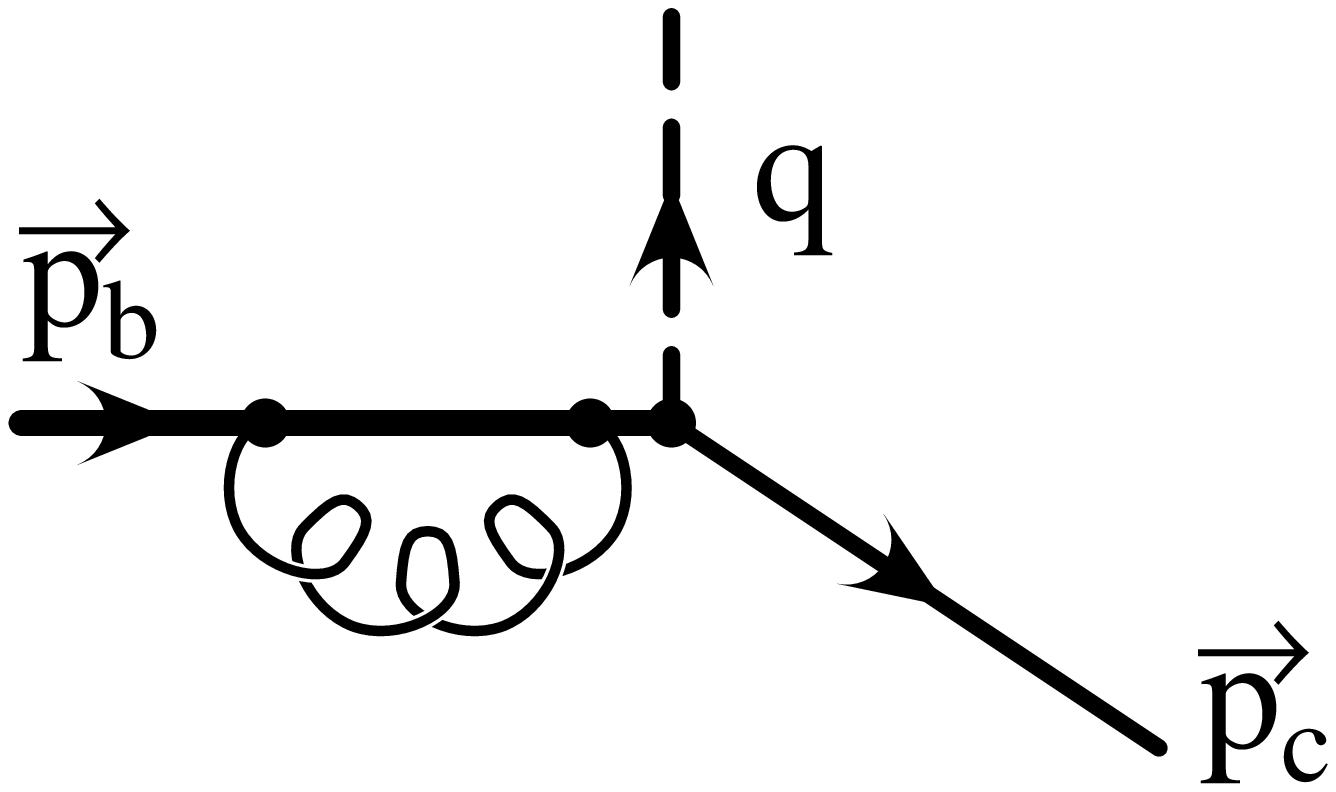,width=50mm,bbllx=160pt,bblly=279pt,%
bburx=530pt,bbury=513pt} 
&   \hspace*{1mm}
\psfig{figure=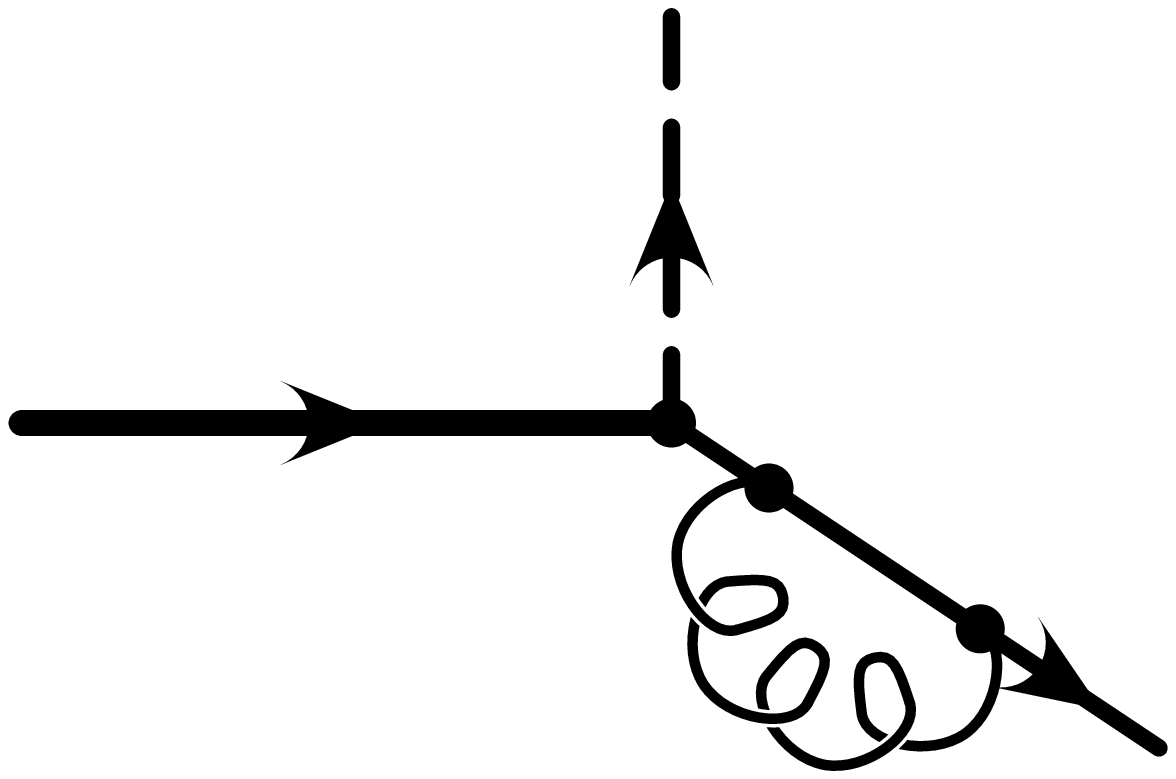,width=50mm,bbllx=160pt,bblly=279pt,%
bburx=530pt,bbury=513pt} 
&\hspace*{1mm}
\psfig{figure=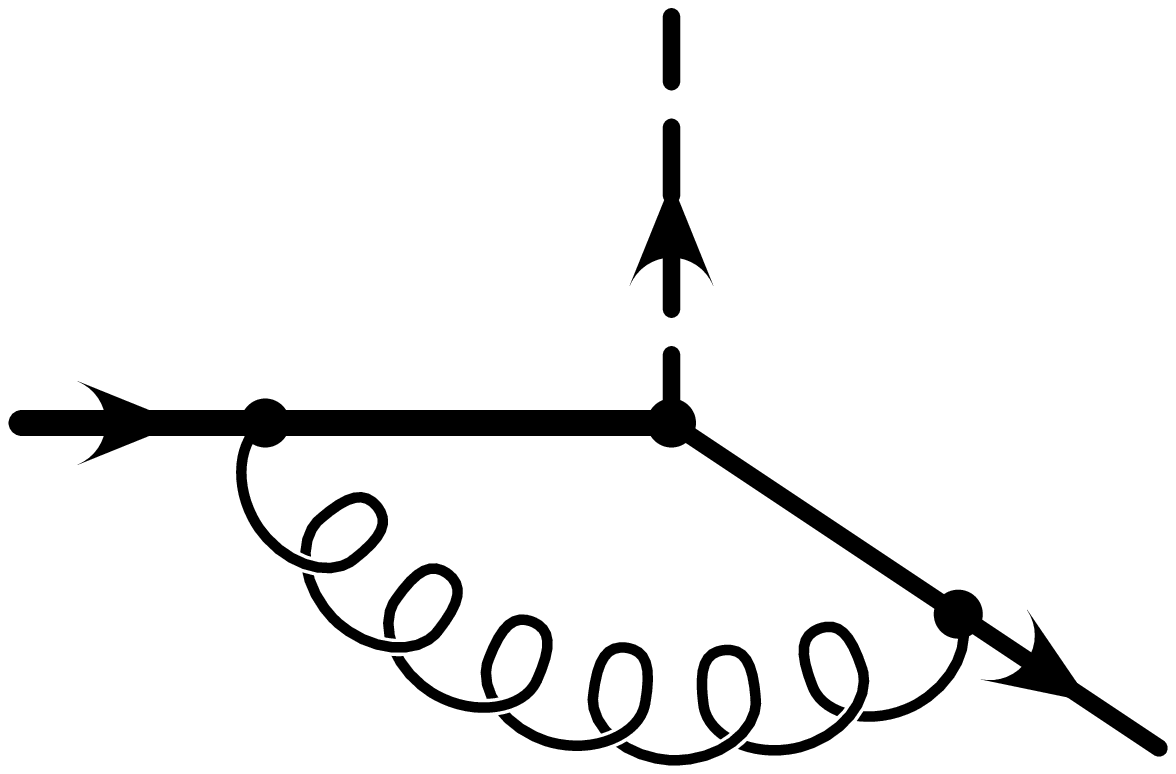,width=50mm,bbllx=160pt,bblly=279pt,%
bburx=530pt,bbury=513pt} 
\end{tabular}}
\]
\end{minipage}
\vspace*{2mm}
\caption{One-loop diagrams determining the
$\alpha_s $-corrections to the nonrelativistic expansion of the vector
current in the small velocity kinematics.}
\label{figs}
\end{figure}

\begin{figure} 
\hspace*{0mm}
\begin{minipage}{16.cm}
\vspace*{3mm}
\[
\hspace*{-19mm}
\mbox{
\psfig{figure=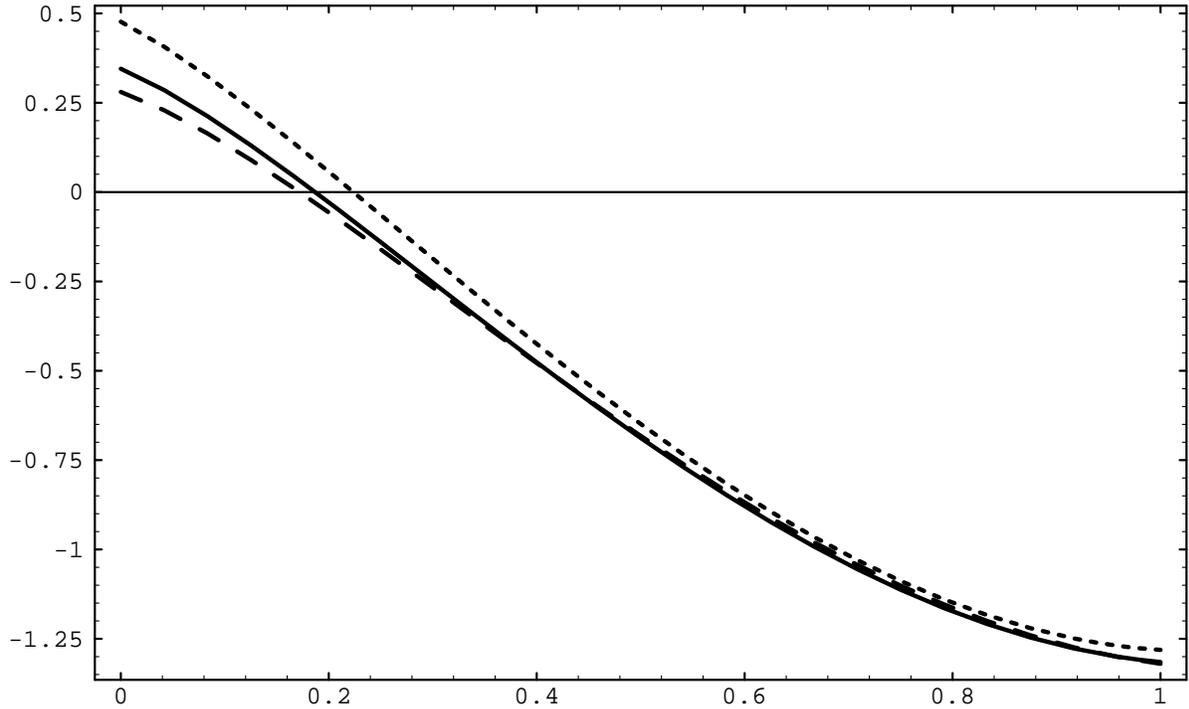,width=100mm,bbllx=150pt,bblly=220pt,%
bburx=450pt,bbury=550pt} 
}
\]
\vspace*{-3mm}
\end{minipage}
\caption{The value of the non-BLM part of the
$\alpha_s ^2$-coefficient $a_2^0$ of the Wilson coefficient $\xi_A(\mu)$ in
the zero recoil axial sum rule for $x=0.2$ (dotted line), $x=0.25$
(solid line) and $x=0.3$ (dashed line) as a function of $\mu/m_c$. The
short-distance renormalization of the $\bar{c} \gamma_k \gamma_5 b$ at
zero recoil is given by $\xi_A^{1/2}$.}
\label{fig:plots}
\end{figure}

\end{document}